\documentclass[a4paper,tightenlines,superscriptaddress,showpacs,nofootinbib,eqsecnum, preprint]{revtex4}

\usepackage[hypertex,a4paper]{hyperref}

\hypersetup{
 colorlinks=false,
 linkcolor=red,
 urlcolor=blue,
 citebordercolor=red,
 linkbordercolor=red,
 pdftitle=EHS Pipeline Paper,
 pdfauthor=Chris Van Den Broeck
}

\usepackage[british]{babel}
\usepackage{float}
\floatstyle{boxed}

\usepackage{graphicx,bm,dcolumn,amsmath,amsfonts,subfigure,color,latexsym,amssymb}

\def\@rcsid{\relax}
\def\rcsid#1{\def\next##1#1{\def\@rcsid{\mbox{RCS ##1}}}\next}

\def\e{\eta}
\def\i{\iota}

\def\ba{\begin{eqnarray}}
\def\ea{\end{eqnarray}}
\def\be{\begin{equation}}
\def\ee{\end{equation}}

\def\nn{\nonumber}

\begin{document}


\draft

\begin{flushright}
gr-qc/0604032
\end{flushright}

\title{Binary black hole detection rates in inspiral gravitational wave searches}
\author{Chris Van Den Broeck}
\email{Chris.van-den-Broeck@astro.cf.ac.uk}

\affiliation{School of Physics and Astronomy, Cardiff University,\\
5 The Parade, Cardiff CF24 3YB, United Kingdom}

\begin{abstract}

The signal-to-noise ratios (SNRs) for quasi-circular binary black hole inspirals computed from restricted post-Newtonian waveforms are compared with those attained by more complete post-Newtonian signals, which are superpositions of amplitude-corrected harmonics of the orbital phase. It is shown that if one were to use the best available amplitude-corrected waveforms for detection templates, one should expect SNRs in actual searches to be significantly lower than those suggested by simulations based purely on restricted waveforms.

\end{abstract}


\maketitle

\section{Introduction and overview}

Gravitational waves from the quasi-circular, adiabatic inspiral of compact binaries consisting of neutron stars and/or black holes are expected to be linear combinations of harmonics of the orbital phase, with the second harmonic as the dominant one. Such waveforms have been computed in the post-Newtonian (PN) approximation, where amplitudes and phases are expressed as expansions in the orbital velocity $v$ (see \cite{Blanchet} for a review and extensive references). For binaries without spin, the best PN waveforms currently available are of order $v^5$ in amplitude \cite{2.5PN} and $v^7$ in phase \cite{3.5PN}. In the usual notation this corresponds to 2.5PN and 3.5PN order, respectively, and we will refer to such waveforms as being of $(2.5,3.5)$PN order. Often the PN corrections to the amplitudes are discarded, in which case one ends up with a single harmonic at twice the orbital phase. This is known as the restricted post-Newtonian waveform. 

Searches for gravitational wave signals from inspiraling compact binaries are performed by matched filtering with a bank of templates. The same kinds of waveforms that go into template banks are also used as simulated signals injected into stretches of data to evaluate algorithms that search for real events and veto spurious ones.  
Several different types of waveforms are in use, all centered on the restricted post-Newtonian approximation. Apart from the straightforward restricted PN waveforms there are the Pad\'e \cite{comparison} and effective one-body \cite{EOB} waveforms, which result from resummation schemes designed to improve on the convergence of the PN phasing; these have no amplitude corrections either. Recently there has been much interest in the phenomenological templates proposed by Buonanno, Chen, Pan, and Vallisneri \cite{BCV}, which do have some simple corrections to the amplitude. Nevertheless, overall there has been a tendency to focus on phasing rather than amplitude, mainly because of the expectation that in matched filtering, the main issue is to know the phasing and number of cycles, or more precisely the number of useful cycles \cite{UsefulCycles}, of the signal in the detector's bandwidth.

Here we investigate the effects of amplitude corrections at high PN order, or the absence thereof, on the reliability of templates and/or simulated signals. Our goal is to arrive at general statements concerning signal-to-noise ratios and detection rates. We will take both restricted and non-restricted waveforms to be the ``standard" ones, so that the waveforms in one family are simple truncations of waveforms in the other, and the two families are parametrized analogously. Even so, when comparing waveforms belonging to different families in the context of detection and parameter estimation, details regarding whether one compares faithfulness as above or effectualness are crucial \cite{comparison,EffFaith}. In some of our considerations these are neglected; however, our main result will be largely independent of them.

Given any two waveforms $h_1$, $h_2$, the signal-to-noise ratio (SNR) for the ``detection" of $h_2$ using (the normalized counterpart of) $h_1$ as a template is given by
\be
\rho_{h_1}[h_2] \equiv \frac{(h_1|h_2)}{\sqrt{(h_1|h_1)}},
\ee
where $(\,.\,|\,.\,)$ is the usual inner product in terms of the noise power spectral density $S_h(f)$ of the detector. With the convention $\tilde{x}(f) = \int_{-\infty}^{\infty} x(t)\,\exp(-2\pi i f t)\,dt$ for Fourier transforms,
\be
(x|y)  \equiv 4\int_{f_{min}}^{f_{max}} \frac{\mbox{Re}[\tilde{x}^\ast (f)\tilde{y}(f)]}{S_h(f)}df.
\label{innerproduct}
\ee

The SNR for ``detecting" a restricted PN waveform $h_0$ using the same waveform for a template would be $\rho_{h_0}[h_0]$, and this is the kind of SNR one encounters in \emph{past and current simulated searches}. When using a restricted template bank in \emph{real} data searches, one may expect the SNR for a genuine signal to be better approximated by $\rho_{h_0}[h]$, with $h$ an amplitude-corrected waveform. For simplicity, assume the parameter values of $h$ and $h_0$ to be identical. As we will demonstrate, one then has
\be
\rho_{h_0}[h] < \rho_{h_0}[h_0],
\label{ineq1}
\ee
the difference being as large as 30\% for astrophysically relevant sources. We stress that in the real detection problem one would maximize the SNR not just over extrinsic parameters but over intrinsic parameters as well. Thus the comparison of faithfulness in Eq.~(\ref{ineq1}) is not adequate and one needs a comparison of effectualness \cite{comparison,EffFaith}. To draw a firm conclusion further study is needed.

However, the main focus will be on a different issue. The inequality (\ref{ineq1}) refers to past and current searches. In \emph{future} searches one might switch to amplitude-corrected templates. As we will show, one has $\rho_h[h] \gtrsim \rho_{h_0}[h]$ with the two differing by only a small amount. Combining this with (\ref{ineq1}), one arrives at
\be
\rho_h[h] < \rho_{h_0}[h_0]
\label{ineq2}
\ee
with the two differing by up to 25\%. Hence, when using the best available amplitude-corrected waveforms for detection templates, one should expect SNRs in actual searches to be significantly lower than those suggested by simulations based purely on restricted waveforms. Since the associated detection rates are proportional to the cubes of the SNRs, they will differ by up to a factor of two. Note that unlike in the interpretation of (\ref{ineq1}), no caveats are needed here (see the analytical treatment in Sec.~\ref{ss:analytic}) beyond the obvious fact that a genuine signal will still differ from a $(2.5,3.5)$PN waveform.

In what follows we first outline the construction of amplitude-corrected waveforms in the stationary phase approximation. Next the above statements concerning SNRs are demonstrated numerically and explained analytically, after which we present conclusions.

For concreteness we will work with the Initial LIGO design sensitivity from~\cite{comparison}, but our results carry over to other initial detectors such as VIRGO.
We set $G = c =1$ unless stated otherwise. A waveform that is of PN order $p$ in amplitude and $q$ in phase will be called a $(p,q)$PN waveform.

\section{The amplitude-corrected waveforms}

The waveforms in the two polarizations take the general form
\be
h_{+,\times}=\frac{2M\eta}{r} x \,
\left\{H^{(0)}_{+,\times} + x^{1/2}H^{(1/2)}_{+,\times} + x H^{(1)}_{+,\times}
+ x^{3/2}H^{(3/2)}_{+,\times} + x^2 H^{(2)}_{+,\times} + x^{5/2} H^{(5/2)}_{+,\times}
\right\} 
\label{hpluscross}
\ee
where $r$ is the distance to the binary, $M$ its total mass, and $\e$ the ratio of reduced mass to total mass. The post-Newtonian expansion parameter is defined as $x= (2\pi M F)^{2/3} = v^2$, with $F(t)$ the instantaneous orbital frequency. The coefficients $H^{(p/2)}_{+,\times}$, $p=0, \ldots, 5$, are linear combinations of various harmonics of the orbital phase with prefactors that depend on the inclination angle $\i$ of the angular momentum of the binary with respect to the line of sight as well as on $\e$; their explicit expressions can be found in \cite{2.5PN}. The measured signal also depends on the polarization angle $\psi$ and the position in the sky $(\theta,\phi)$ through the detector's beam pattern functions $F_{+,\times}$:
\be
h(t)=F_+ h_+(t) + F_\times h_\times(t). \label{signal}
\ee
For ground-based detectors, which are the ones we will be concerned with, it is reasonable to approximate $F_{+,\times}$ as being constant in time for the duration of the signal in the detector's bandwidth. The signal (\ref{signal}) is a linear combination of harmonics of the orbital phase $\Psi(t)$ with offsets $\varphi_{(k,m/2)}$. We consider $(2.5,3.5)$PN waveforms, which contain seven harmonics:
\be
h(t) = \sum_{k=1}^7 \sum_{m=0}^5 A_{(k,m/2)}(t) \cos(k\Psi(t) + \varphi_{(k,m/2)}). \label{sum}
\ee
The index $k$ runs over the harmonics while $m/2$ is PN order in amplitude. The harmonic at twice the orbital phase dominates, as it is the only one with a 0PN amplitude contribution.

In the \emph{restricted} post-Newtonian approximation, no amplitude corrections are taken into account. In that case only one harmonic is present, namely the one at twice the orbital phase:
\be
h_0(t) = A_{(2,0)}(t) \cos(2\Psi(t) + \varphi_{(2,0)}).
\ee

During the adiabatic part of the inspiral one has $|d\ln A_{(k,s)}/dt| \ll k d\Psi/dt$ and 
$|k d^2\Psi/dt^2| \ll (k d\Psi/dt)^2$ for $k=1,2,\ldots$, in which case one can use the stationary phase approximation (SPA) to the Fourier transform of (\ref{sum}):
\be
\tilde{h}_{SPA}(f) = \sum_{k=1}^7 \left[\frac{\sum_{m=0}^5 A_{(k,m/2)}\left(t\left(\frac{1}{k}f\right)\right)\,e^{-i\varphi_{(k,m/2)}}}{2\sqrt{k \dot{F}\left(t\left(\frac{1}{k} f\right)\right)}}\right]_{2.5} 
\exp\left[i\left(2\pi f t_c -\pi/4 + k \psi\left(\frac{1}{k} f\right)\right)\right], 
\label{SPA}
\ee
where $[\,.\,]_{2.5}$ denotes consistent truncation to $2.5$th post-Newtonian order (i.e., the ``Newtonian" prefactor $f^{-7/6}$ is taken outside and the remaining expression is expanded in $(2\pi M f)^{1/3}$ up to fifth power).
A dot denotes derivation with respect to time and $t_c$ is the coalescence time. The function $t(f)$ is defined implicitly through the instantaneous orbital frequency by $F(t(f)) = f$. The ``frequency sweep" $\dot{F}$ in terms of $F$ can be obtained from the expressions for energy and flux in \cite{3.5PN}, and the same goes for the phase $\psi(f)$.

Finally, the expressions of the respective harmonics are only valid below some upper cut-off frequency. The $k$th harmonic will be terminated at a frequency $k f_{LSO}$, where $f_{LSO}$ is the orbital frequency corresponding to the last stable orbit. In the time domain this roughly corresponds to cutting off \emph{all} of the harmonics at a time determined by $F(t) = f_{LSO}$. In the point mass limit one has
$f_{LSO} = (6^{3/2} 2\pi M)^{-1}$,
and for simplicity this is the expression we will adopt. To impose these restrictions on the harmonics, in practice we multiply the $k$th harmonic by $\theta(k f_{LSO}- f)$, where $\theta(x)$ is the usual Heaviside step function. In the definition of the inner product $(\,.\,|\,.\,)$, Eq.~(\ref{innerproduct}), it is then natural to take $f_{max} = 7 f_{LSO}$, the frequency reach of our highest harmonic. The lower cut-off frequency $f_{min}$ is detector-dependent; for Initial LIGO it is set to 40 Hz.

The SPA is just one method of approximating the Fourier transform; the main alternative is the fast Fourier transform (FFT). As shown in \cite{SPAvsFFT,UsefulCycles} in the context of Initial LIGO, SNRs for restricted waveforms computed using the FFT and SPA differ by only a few percent for total masses up to $\sim 30\,M_\odot$, and this is the mass range we will focus on. We also note that the conditions for the applicability of the SPA on the harmonics become more favorable with increasing $k$.

\section{Signal-to-noise ratios}

\subsection{Numerical observations}

To set the stage, let us look at the SNRs in Initial LIGO for three different inspiral events at a distance of 20 Mpc: a binary neutron star (NS--NS), a neutron star and a black hole (NS--BH), and a binary black hole (BH--BH). The mass of the neutron star is taken to be $1.4\,M_\odot$ while for the black hole we choose $10\,M_\odot$. Table \ref{t:PNdependence} displays the variation of $\rho_{h_0}[h_0]$, $\rho_{h_0}[h]$, and $\rho_h[h]$ for $(p,3.5)$PN waveforms $h$ with increasing $p$, where $h_0$ is the corresponding restricted waveform. The most interesting case is the NS--BH system. In going from $p=0$ to $p=0.5$, both $\rho_{h_0}[h]$ and $\rho_h[h]$ increase, as has been noted for $\rho_h[h]$ by Sintes and Vecchio \cite{SV1}. However, at $p=1$ they drop below $\rho_{h_0}[h_0]$, and a downward trend is seen as $p$ continues further. At $p=2.5$ one has $\rho_{h_0}[h]<\rho_{h_0}[h_0]$, with $\rho_{h_0}[h_0]$ being larger by 16.5\% of $\rho_{h_0}[h]$ -- modeling signals as restricted waveforms and searching for them with restricted templates leads to an overestimation of SNR. Still at 2.5PN, $\rho_h[h] > \rho_{h_0}[h]$ but with $\rho_h[h]$ being larger by only 3.6\% of $\rho_{h_0}[h]$; hence there is only a modest improvement to be had from the use of amplitude-corrected templates. Specifically, $\rho_h[h] < \rho_{h_0}[h_0]$ with a difference of 12.4\%. The NS--NS and BH--BH systems exhibit similar trends.

\begin{table}[htp!] 
\begin{tabular}{lrcccrcccrccc} 
\hline 
\hline 
&\vline& & NS--NS & &\vline& & NS--BH & &\vline& & BH--BH & \\
\hline
$p$ &\vline& \,\,\,$\rho_{h_0}[h_0]$\,\,\, & \,\,\,$\rho_{h_0}[h]$\,\,\,  & \,\,\,$\rho_h[h]$ &\vline& 
\,\,\,$\rho_{h_0}[h_0]$\,\,\, & \,\,\,$\rho_{h_0}[h]$\,\,\,  & \,\,\,$\rho_h[h]$ &\vline& 
\,\,\,$\rho_{h_0}[h_0]$\,\,\, & \,\,\,$\rho_{h_0}[h]$\,\,\,  & \,\,\,$\rho_h[h]$\\
\hline 
0   &\vline& 6.465 & 6.465 & 6.465 &\vline& 13.492 & 13.492 & 13.492 &\vline& 30.928 & 30.928 & 30.928 \\
0.5 &\vline& "     & 6.465 & 6.465 &\vline&  "     & 13.503 & 13.932 &\vline& "      & 30.928 & 30.928 \\ 
1   &\vline& "     & 6.285 & 6.286 &\vline&  "     & 12.048 & 12.563 &\vline& "      & 28.108 & 28.135 \\
1.5 &\vline& "     & 6.285 & 6.286 &\vline&  "     & 12.049 & 12.421 &\vline& "      & 28.108 & 28.135 \\
2   &\vline& "     & 6.247 & 6.249 &\vline&  "     & 11.669 & 12.090 &\vline& "      & 26.319 & 26.373 \\
2.5 &\vline& "     & 6.246 & 6.247 &\vline&  "     & 11.584 & 12.002 &\vline& "      & 26.217 & 26.285 \\
\hline
\end{tabular}
\caption{Change in signal-to-noise ratios with increasing $p$ in $(p,3.5)$PN waveforms, for three different systems at a distance of 20 Mpc, as seen in Initial LIGO. (Angles were chosen arbitrarily as $\theta=\phi=\pi/6$, $\psi=\pi/4$, $\i=\pi/3$.)} 
\label{t:PNdependence}
\end{table}

In the Table we chose specific values for the angles in the problem. Once again consider the NS--BH system and let $h$ be a $(2.5,3.5)$PN waveform. Then for an optimally located and oriented source ($\theta=\iota=0$), $\rho_{h_0}[h_0]$ is larger than $\rho_{h_0}[h]$ by 12.4\%, but $\rho_h[h]$ is larger than $\rho_{h_0}[h]$ by only 0.14\%. For a rather inconveniently located and oriented source (say, $\theta = \pi/2$, $\phi = \psi = 0$, and $\iota = \pi/2$), the corresponding numbers are 19.2\% and 5.1\%, respectively. As these examples indicate, the relative difference between $\rho_h[h]$ and $\rho_{h_0}[h_0]$ has only a weak angular dependence, varying between 12\% and 14\% for the system at hand.

\begin{figure}[htbp!]
\centering
\includegraphics[scale=0.40,angle=0]{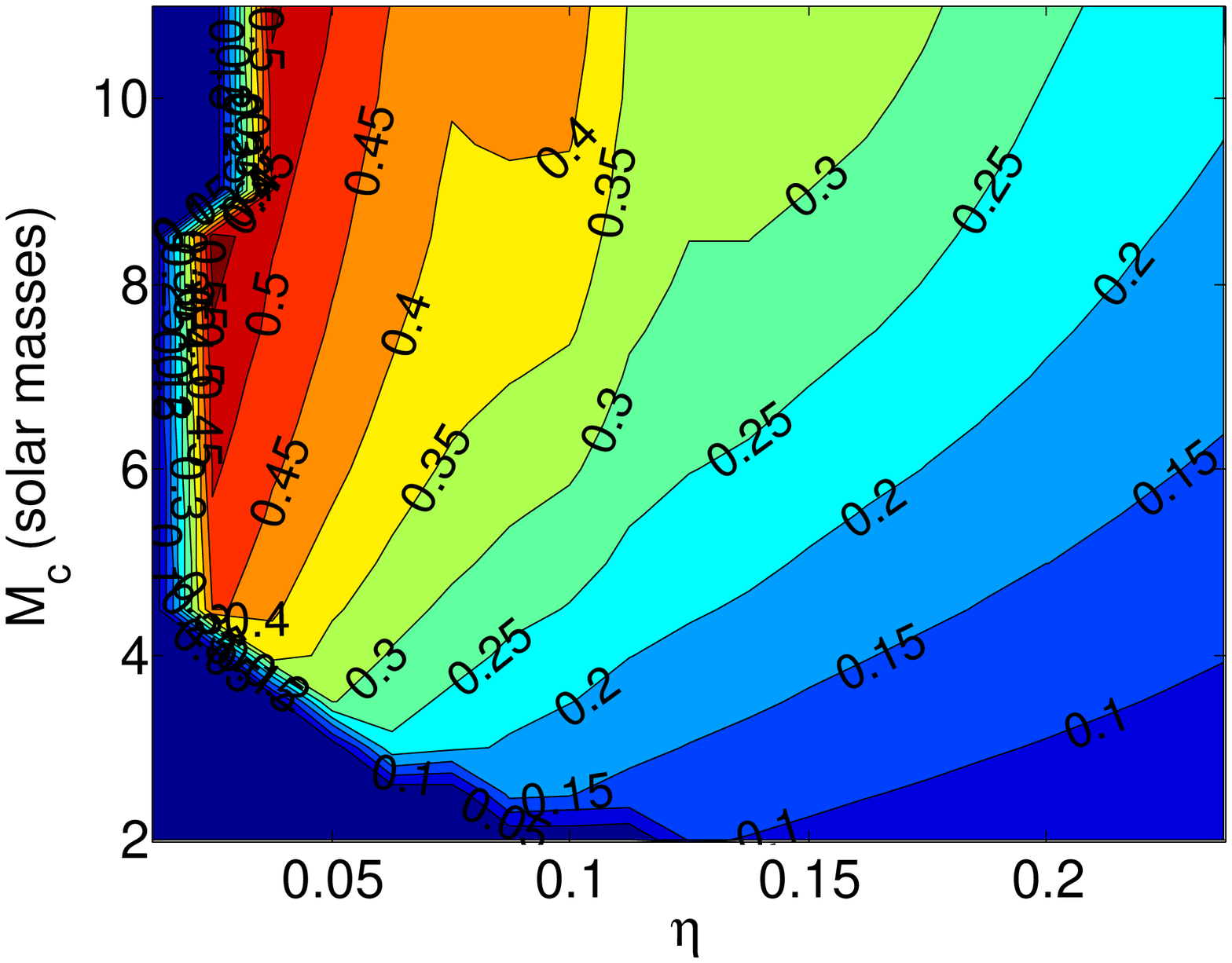}
\includegraphics[scale=0.40,angle=0]{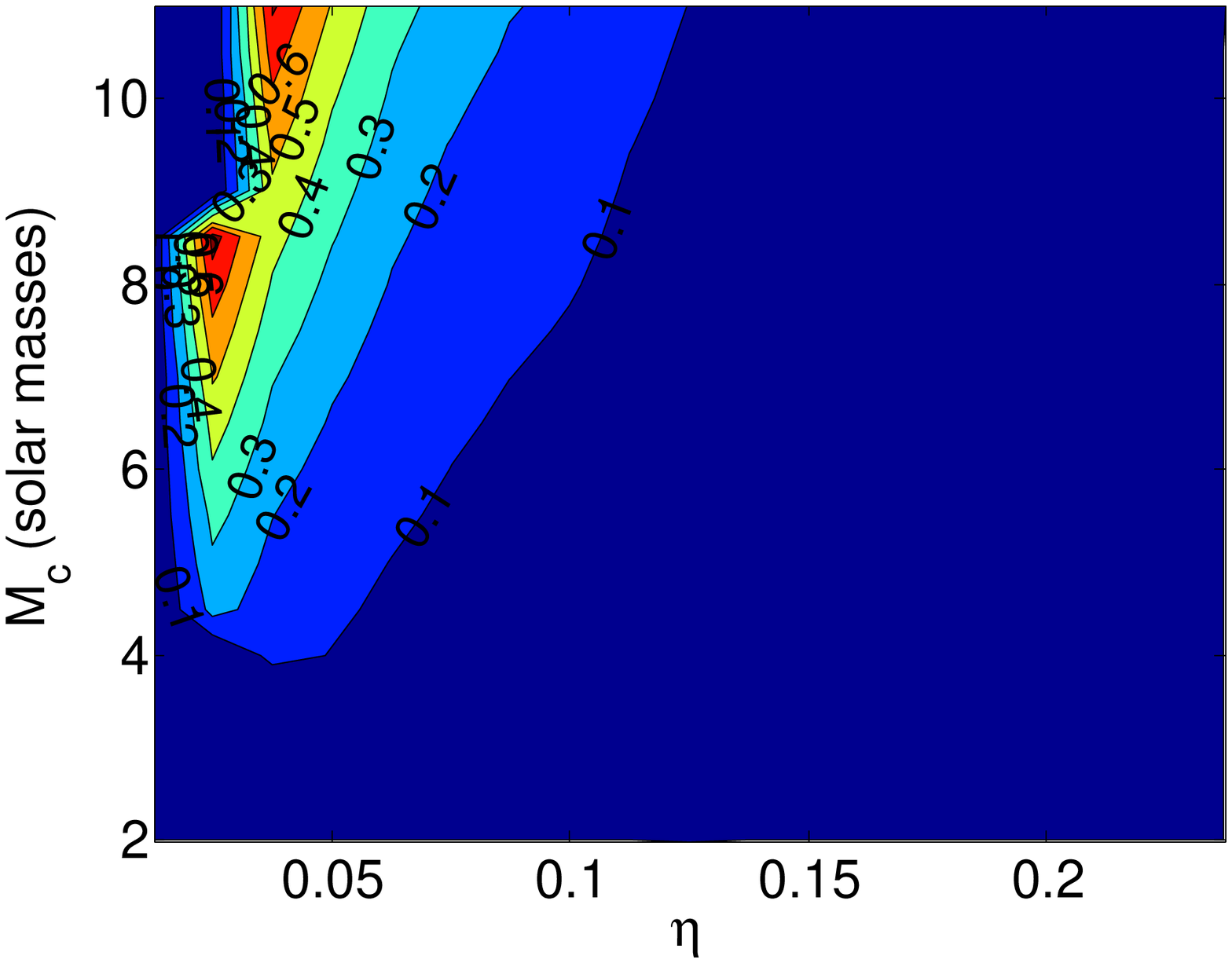}
\caption{Plots of $\delta$ (left panel) and $\gamma$ (right) as functions of $\eta$ and chirp mass $M_c=M\eta^{3/5}$. In both plots we have imposed that individual component masses be at least 1 $M_\odot$ and that total mass $M < 80\,M_\odot$; hence the sudden drop on the left in each panel.}
\label{f:rhoDvsrho}
\end{figure}  

This behavior is quite generic. In the left panel of Fig.~\ref{f:rhoDvsrho} we compare $\rho_{h_0}[h]$, where $h$ is a $(2.5,3.5)$PN waveform, with $\rho_{h_0}[h_0]$. The fraction by which SNR is being overestimated by modeling signals as restricted waveforms is
$\delta \equiv (\rho_{h_0}[h_0] - \rho_{h_0}[h])/\rho_{h_0}[h]$.
In a significant part of parameter space this fraction is between 20 and 30\%. Indeed, on the line corresponding to $\delta=0.2$, lighter and heavier component masses $m_1$, $m_2$ are constrained to the ranges $1\,M_\odot \lesssim m_1 \lesssim 12\,M_\odot$ and $12\,M_\odot \lesssim m_2 \lesssim 15\,M_\odot$, while for $\delta=0.3$ one has $1\,M_\odot \lesssim m_1 \lesssim 8\,M_\odot$ and  $17\,M_\odot \lesssim m_2 \lesssim 24\,M_\odot$. The implied range of binary systems is of clear astrophysical interest. As already suggested by the results of Table \ref{t:PNdependence}, the overestimation is more pronounced for asymmetric systems and shows a strong dependence on mass.

Next, let us look at the gain in SNR obtained by using amplitude-corrected waveforms, rather than restricted ones, as templates, with the signal being an amplitude-corrected waveform. The fractional gain is
$\gamma \equiv (\rho_h[h] - \rho_{h_0}[h])/\rho_{h_0}[h]$.
The right panel of Fig.~\ref{f:rhoDvsrho} shows a plot of $\gamma$ as a function of chirp mass and $\eta$. Clearly the use of non-restricted waveforms as templates would not give us much advantage over restricted ones; in most of parameter space the difference between the two is in the order of a few percent. Large gains ($> 20$\%) do occur, but only for the astrophysically less interesting cases with large $M_c$ and small $\eta$, corresponding to very asymmetric binaries with high total mass ($\gtrsim 40\,M_\odot$). 

Roughly speaking, the left panel in Fig.~\ref{f:rhoDvsrho} indicates by how much SNR has hitherto been overestimated due to inadequate signal modeling. The right panel shows that, even if in future searches one were to use amplitude-corrected waveforms throughout, this overestimation will not get compensated for. Indeed, putting together the information from both panels, we may conclude that
\be
\rho_h[h] < \rho_{h_0}[h_0]
\label{mainresult}
\ee
with the two sides differing by up to 25\% for sources with $M \lesssim 30\,M_\odot$. 

\subsection{Analytic considerations}
\label{ss:analytic}

In this subsection, for brevity we focus on the inequality
\be
\rho_{h_0}[h] < \rho_{h_0}[h_0],
\label{ineq} 
\ee
but its analytic explanation will already provide some insight into the more interesting result (\ref{mainresult}); a much more detailed analysis will be given in \cite{inpreparation}. The above inequality is equivalent to $(h_0|h) < (h_0|h_0)$. Writing $h$ as a sum of amplitude-corrected harmonics $h^{(k)}$, the LHS may be approximated as
\be
(h_0|h) = \sum_{k=1}^7 (h_0|h^{(k)}) 
        \simeq (h_0|h^{(2)}), 
\label{approximation} 
\ee
because different harmonics tend to interfere destructively. (It can be checked numerically that the approximation is valid to within a few percent in the Initial LIGO band.) We stress that $h^{(2)} \neq h_0$. Indeed, both are proportional to the second harmonic, but $h^{(2)}$ contains amplitude corrections while $h_0$ does not.

The SPA for $h^{(2)}$ is of the form (leaving out the step function in frequency):
\be
\tilde{h}^{(2)}(f) = \mathcal{C} f^{-7/6} 
\left[ \sum_{m=0}^5 C_{(m/2)} (2\pi M f)^{m/3} \right]
\exp\left[i\left(2\pi f t_c -\pi/4 + 2\psi\left(\frac{1}{2} f\right)\right)\right],
\label{h2}
\ee
where $\mathcal{C}$ is a real function of chirp mass and distance while the coefficients $C_{(m/2)}$ are complex functions of $(\theta,\phi,\psi,\iota,\e)$. Setting $C_{(m/2)} = 0$ for $m \geq 1$ would lead to the restricted waveform $h_0$. 

Consider the form of the quantity $\mbox{Re}[\tilde{h}^\ast_0(f) \tilde{h}^{(2)}(f)]$ appearing in the integrand of $(h_0|h^{(2)})$:
\ba
\mbox{Re}[\tilde{h}^\ast_0(f) \tilde{h}^{(2)}(f)] = \mathcal{C}^2 f^{-7/3}
\sum_{m=0}^5 \mbox{Re}[C^\ast_{(0)} C_{(m/2)}] (2\pi M f)^{m/3}.
\label{h0h2}
\ea
This is to be compared with the expression $\mbox{Re}[\tilde{h}^\ast_0(f) \tilde{h}_0(f)] = \mathcal{C}^2 f^{-7/3} |C_{(0)}|^2$ in the integrand of $(h_0|h_0)$. Now, it is not difficult to show that
\be
\mbox{Re}[C^\ast_{(0)} C_{(m/2)}] \leq 0
\label{inequality1}
\ee
for $m = 1, \ldots, 5$, \emph{irrespective of parameter values}. In particular, the functions $\mbox{Re}[C^\ast_{(0)}C_{(1/2)}]$ and $\mbox{Re}[C^\ast_{(0)}C_{(3/2)}]$ vanish identically while the others are negative definite. Consequently, 
\be
\mbox{Re}[\tilde{h}^\ast_0(f) \tilde{h}^{(2)}(f)] < \mbox{Re}[\tilde{h}^\ast_0(f) \tilde{h}_0(f)].
\ee
Thus, one should expect $(h_0|h) < (h_0|h_0)$, whence $\rho_{h_0}[h] < \rho_{h_0}[h_0]$.

Next we revisit Table \ref{t:PNdependence}. Let $h$ be a $(p,3.5)$PN waveform with $0 < p \leq 2.5$.
If $p=0.5$, only the first term in the RHS of (\ref{h0h2}) survives \cite{SV1}. In that case $(h_0|h) \gtrsim (h_0|h_0)$ due to the small contributions $(h_0|h^{(1)})$ and $(h_0|h^{(3)})$ to the LHS, which were neglected in (\ref{approximation}). However, as $p$ is increased there is a notable downward trend in $\rho_{h_0}[h]$ as a result of the inequalities (\ref{inequality1}), except at $p=1.5$ because of $\mbox{Re}[C^\ast_{(0)}C_{(3/2)}]=0$.

\begin{table}[htp!] 
\begin{tabular}{crcc} 
\hline 
\hline 
&\vline& $\rho_{h_0}[h]$ & $\rho_{h_0}[h^{(2)}]$ \\
\hline
NS--NS &\vline& 6.2462 & 6.2462 \\ 
NS--BH &\vline& 11.5836 & 11.5731 \\ 
BH--BH &\vline& 26.2166 & 26.2126 \\
\hline
\end{tabular}
\caption{Signal-to-noise ratios for detection of $(2.5,3.5)$PN signals $h$ and the amplitude-corrected second harmonic $h^{(2)}$, using restricted templates, for three ``canonical" systems in Initial LIGO at a distance of 20 Mpc.} 
\label{t:Hybrid}
\end{table}

The above analysis suggests that when using restricted templates it should be possible to construct adequate signal waveforms without having to include higher harmonics. Indeed, it should suffice to reinstate the amplitude corrections to the second harmonic only, in which case we arrive at the waveforms $h^{(2)}$. This is borne out in Table \ref{t:Hybrid}.

Although slightly more involved, the inequality $\rho_h[h] < \rho_{h_0}[h_0]$ can also be proved analytically by continuing along the lines above \cite{inpreparation}. This direct proof not going via $\rho_{h_0}[h]$ is more explicitly free of the issue of maximization over intrinsic parameters.

\section{Conclusions}

We have compared the performance of the best available amplitude-corrected post-Newtonian waveforms, namely the $(2.5,3.5)$PN waveforms, with that of the restricted ones, both as templates and as simulated signals. Our results can be summarized as follows.

(i) In simulated searches, templates and simulated signals alike have hitherto been modeled essentially as restricted waveforms. In real searches, a genuine signal will presumably be better approximated by an amplitude-corrected waveform, and we have seen that the associated SNRs may be considerably smaller than those suggested by the simulations. However, issues concerning maximization over parameters were neglected; it would be of interest to perform a more in-depth study.

(ii) More importantly, in the future one might use amplitude-corrected waveforms both for templates and simulated signals. In that case one may expect simulated SNRs to be much closer to the ones seen in actual searches conducted using the amplitude-corrected templates. We found that SNRs computed with amplitude-corrected templates and waveforms are smaller than the ones with restricted templates and waveforms by up to 25\% for astrophysically interesting sources, corresponding to a difference in detection rates by up to factor of two. This result is much more robust; except for the fact that a genuine signal will still differ from a $(2.5,3.5)$PN waveform, it is independent of any other detection issues. It implies that if one were to use the best available amplitude-corrected waveforms for detection templates, one should expect SNRs in actual searches to be significantly lower than those suggested by simulations that involve only restricted waveforms.

The effects seen here result largely from the amplitude corrections to the dominant harmonic. An interesting consequence is that by taking only these into account, one arrives at signal models which, given restricted templates, behave like the more complete waveforms despite the fact that they lack other harmonics. Overall, our results clearly underscore the importance of including amplitude corrections in templates and simulated signals. A more detailed treatment will be given in a forthcoming paper \cite{inpreparation}.

\section*{Acknowledgements}

It is a pleasure to thank S.~Babak, B.S.~Sathyaprakash and A.S.~Sengupta for very useful discussions. I am also grateful to an anonymous referee whose suggestions greatly improved the presentation. This research was supported by PPARC grant PP/B500731/1.

\end{document}